# Toward New Vision in Teaching Calculus

Seifedine Kadry[a,*], Maha El Shalkamy[b]

[a]*American University of the Middle East, Kuwait*
[b]*American university of Cairo, Egypt*

**Abstract**

Usually the first course in mathematics is calculus. It's a core course in the curriculum of the Business, Engineering and the Sciences. However many students face difficulties to learn calculus. These difficulties are often caused by the prior fear of mathematics. The students today can't live without using computer technology. The uses of computer for teaching and learning can transform the boring traditional methodology of teach to more active and attractive method. In this paper, we will show how we can use Excel in teaching calculus to improve our students' learning and understanding through different types of applications ranging from Business to Engineering. The effectiveness of the proposed methodology was tested on a random sample of 45 students from different majors over a period of two semesters.

Keywords: teaching methodology, visual approach, Microsoft Excel, calculus, statistical evaluation.

## 1. Introduction

Teaching methods can best be defined as the types of principles and methods used for instruction. There exist different types of teaching methodologies based on the type of information or expertise the instructor is trying to deliver. Power point presentation, class discussion, problem-solving, and memorization are some of the teaching methods being used. When an instructor is choosing a teaching method, he/she needs to adapt their style according to their students. The efficiency of the teaching methods is the major factor to lead the student to succeed [1]. The computer technology is widely used by students and it's become an important part of their daily life. In this paper, we will propose a new visual methodology for teaching instead of the traditional and passive methodology by using computer. We will use Microsoft office excel software to teach calculus course, calculus is a major course for many majors. The experimental test shows the efficient of the propose method. This paper is organized as follows. Section 2 introduces Microsoft Office Excel. Section 3 gives an overview of the existing teaching methods. In the section 4, we will illustrate the using of excel through varieties of calculus topics ranging from business, engineering and sciences. The experimental test and its result are discussed in section 5 and finally the conclusion and the perspective are given in section 6.

* Corresponding author. Tel.: 00965-66610985
 *E-mail address:* skadry@gmail.com.

**2. Why Microsoft Office Excel**

Microsoft office is very essential packages for any computers. Excel is a part of this package. Excel is electronic spread sheet software, that end user use to store and structure their data in a tabular form. By using the spread sheet, we can do any arithmetic application and use a huge set of mathematics built in function in addition to build our formula [7]. It features calculation, draw graphics, create formulas, and solve equations and many mathematical functions. The usage of excel is easy and straightforward. The recent versions of Excel are 2008 for Mac and 2010 for Windows.

**3. Teaching Methods in Education**

For effective teaching to take place, a good method must be adopted by a teacher. In this section we will describe briefly some of well-known teaching methods such as the lecture method, the cooperative method, the demonstration method, the discussion method, and finally computer-based method [2, 7].

*Lecture Method*
The traditional and broadly used method is the lecture method. Lecture method takes the form of presentation. The instructors should know how to organize and present the lecture. Moreover, they must understand the drawbacks and the advantages of this method. This method is used to introduce new topics, present graphical subjects, summarize ideas, present coding, and display the main points of the lecture, showing the relation between the theory and the practice. To add some interactivity, we can combine the lecture method with other teaching methods like the discussion method.

*Cooperative Method*
An attractive strategy of teaching is to have the students working together in groups during the session. This is the concept of the cooperative method. Several researches on this method indicate: the increasing of their participation, improve their communication, augmenting their knowledge, encouraging their development of the social skill, increasing their independence and simply facilitates their effective learning. In a cooperative method, one classroom may share by more than one instructor and these instructors will share all responsibility including managing groups, planning, coaching, tutoring and grading. In this method, it's important to take into consideration the subsequent points: the size of each group, the selection and the distribution of each task and the choice of each student in its appropriate group.

*Demonstration Method*
The concept of this method is "learning by doing". This method of teaching supplies the students a comprehensible image of the subject that must be learned. This method is suitable for teaching a skill because it covers all the necessary steps in an effective learning order. It refers to the capability of students to improve their skills by regularly repeating the same type of action. An individual learns to calculate by calculating, to write by writing, to swim by swimming, and to drive by driving. This method allows the students to relate the principles and theories to a practical situation. This method is more relies on the student and consequences of an effective student participation and involvement than any other teaching method. The demonstration method requires a high degree of instructor skill and it's limited to small group of students.

*Discussion Method*

This method is very active contrary to the passive lecture method. Here the instructor presents information for only a short period of time and query the student then engages them in open discussion for a while on particular issues related to the topic. This not only helps students participate more actively, it can also reveal to the instructor whether the students generally understand and relate to the material being presented. The target of the instructor is to draw out what the students understand, rather than to spend the class period telling them. The larger participation and the dynamic discussion lead to a more efficiency in the learning process. All students in the group should contribute in the discussion and everyone should feel himself/herself as a part of this discussion. The instructor should take care of everyone in fairly manner and should give them confidence, encourage them to ask questions and to answers. Cynicism should never be used, since it inhibits the concentration of the participants and reduce the seriousity. Finally, the session achieves closure when the lecturer summarizes and consolidates main points from the lecture and discussion.

*Computer-Based Method*

This method is delivering instructional content and activities to students via computers. Here, computers are tools which complete and strengthen the system; they are not alternatives which replace teachers in teaching process. In computer based method, computers are used to support education and instruction. Classroom instructor is the main entity that teaches the subject, and determined objectives and attitudes. In this method, an instructor can use computers in different periods, places and ways while teaching according to the characteristics of the students and the type of the subjects that will be teach.

**4. Learn some Calculus Topics using Excel**

In this section, we will illustrate the using of excel through selected topics from calculus course.

*Case 1: optimization problem.*

The total cost of producing x products is $C(x) = 42x + 16800/x$. our goal is to search the value of x, where the cost function $C(x)$ is minimum. Step 1: First graph $C(x)$ by making a table and invoking ChartWizard (fig. 2):

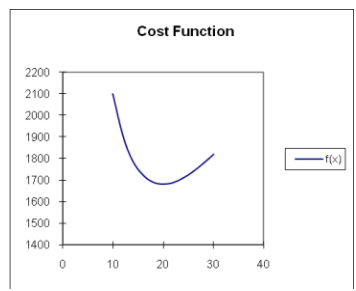

Figure 2: case 1

Using the above graph, the minimum approximately is be between 10 and 30. Now use Goal Seek with $f'(x) = 42-16800/x^2$. Step 2: Since $f'(x) =0$ when the function has a maximum or minimum, we can use Goal Seek to figure out when $f'(x) =0$. Set initial value to 3 since minimum occurs close to 3 (table 3):

| x | f'(x) |
|---|---|
| 3 | -12 |

Table 3: value for case 1

From the Tools menu, select the Goal Seek... option, you will see a dialog box. You want to vary x so that the f'(x). Now, the true value of x should displays in the above table.

P.S.: The minimum and the maximum lead to f'(x) = 0. Generally, Goal Seek gave the solution closest to initial value of x (3 in this example). For this reason is better to sketch the graph.

*Case 2: calculation of the interest.*

In this example, we shall explore how compounded interest is calculated. To illustrate the process, we will produce tables to get a perceptive feel. Example: Suppose you earn 4% interest, compounded quarterly, on a bank account. You deposit an amount of $100 in the bank. How much interest do you accrue provided you do not withdraw any money?, first we make a table (table 4) as follows:

| Date | Deposits | Interest | Balance |
|---|---|---|---|
| 1/1/1994 | $100.00 | | $100.00 |
| 4/1/1994 | | $ 1.00 | $101.00 |
| 7/1/1994 | | | |
| 10/1/1994 | | | |
| 1/1/1995 | | | |

Table 4: values for case 2

Note that the interest is calculated by multiplying the previous balance by 1%. Since the interest is compounded quarterly, the interest rate per quarter is annual rate/ # periods = 4%/4 = 1%.

**Case 3**: areas and Riemann sums.

We will show how we can use Excel to approximate the value of a definite integral via Riemann sums. Example: Approximate the area under the curve y= x+2 from x=1 to x=3 using 10 equal subintervals and the right endpoints of the subintervals to construct the rectangles. We construct a table (table 5) as follows - since the interval is 2 units, each subinterval will be of length 0.2. The right endpoints will then be 1.2, 1.4...3:

| x_i | delta_x | f(x_i) | f(x_i)*delta_x |
|---|---|---|---|
| 1.2 | 0.2 | 3.2 | 0.64 |
| 1.4 | 0.2 | 3.4 | 0.68 |
| . | . | . | . |
| . | . | . | . |
| 2.8 | 0.2 | 4.8 | 0.96 |
| 3 | 0.2 | 5 | 1 |
| | | Total | 8.2 |

Table 5: values for case 3

**5. Evaluation of the Experimental Test**

To show the efficiency of the proposed method, we selected randomly 45 students from different majors over a period of two semesters. Following the statistical result (table 6):

| methods | Lecture method | using Excel |
|---|---|---|
| Mean | 66.2 | 80.1 |

Table 6: average grade comparison

A paired t-test was performed to determine if the proposed methodology was effective, the following result (table 7) shows clearly that the proposed method is better than the traditional method:

t-Test: Two-Sample Assuming Unequal Variances

|  | Variable 1 | Variable 2 |
|---|---|---|
| Mean | 65.93478261 | 80.45652174 |
| Variance | 260.5956522 | 304.2980676 |
| Observations | 45 | 45 |
| Hypothesized Mean Difference | 0 | |
| Df | 89 | |
| t Stat | -4.14394682 | |
| P(T<=t) one-tail | 3.88116E-05 | |
| t Critical one-tail | 1.662155326 | |
| P(T<=t) two-tail | 7.76231E-05 | |
| t Critical two-tail | 1.986978657 | |

Table 7: result of T-test

After the t-test, we performed ANOVA test (table 8-9-10) by partitioning the selected students into three groups based on their major. The result shows an efficiently difference between the majors (Business, Engineering and sciences) using the proposed methodology:

| Groups | Count | Sum | Average | Variance |
|---|---|---|---|---|
| Engineering | 15 | 1241 | 82.73 | 238.49 |
| sciences | 15 | 1185 | 79 | 304.42 |
| business | 15 | 1180 | 78.66 | 408.80 |

Table 8: Summary table

| ANOVA | | | |
|---|---|---|---|
| Source of Variation | SS | df | MS |
| Between Groups | 152.93 | 2 | 76.4667 |
| Within Groups | 13324.26 | 42 | 317.2444 |
| Total | 13477.2 | 44 | |

Table 9: ANOVA test result

| F | P-value | F crit |
|---|---|---|
| 0.241033903 | 0.786894473 | 3.219942293 |

Table 10: P-value

The details (table 11), the statistic result (table 12) and the box plot graph (figure 4) between different majors is given in the following:

| Engineering | sciences | business |
|---|---|---|
| 89 | 97 | 61 |
| 60 | 79 | 51 |
| . | . | . |
| . | . | . |
| 82 | 79 | 98 |
| 98 | 97 | 94 |

Table 11: average of each student

| Statistic | Engineering | Sciences | Business |
|---|---|---|---|
| q1 | 75 | 70 | 58.5 |
| min | 46 | 45 | 48 |
| median | 89 | 79 | 87 |
| max | 98 | 99 | 100 |
| q3 | 95 | 95.5 | 95 |

Table 12: statistic result

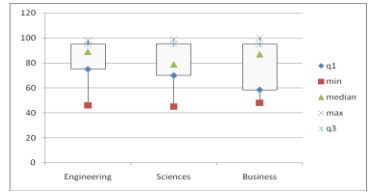

Figure 4: Box plot of different majors

## Conclusion and future works

The visual approach increases the percentage of concentration of the students and especially when the student can produce result in a short time like using Excel. The advantages of using Excel to learn calculus are: understand very well the different topics in calculus in an interesting way, get more experience and knowledge in Excel which will be using it in the future. In the future works, we will test others mathematics software like Matlab, MathCAD and Maple and to compare their efficiency with Excel to a different students majors.